CHALLENGES OF RELATIVISTIC ASTROPHYSICS

Reuven Opher
University of Sao Paulo, Sao Paulo, Brazil
Cornell University, Ithaca, NY, USA

E-mail: opher@astro.iag.usp.br

ABSTRACT

I discuss some of the most outstanding challenges in relativistic astrophysics in the subjects of: COMPACT OBJECTS (Black Holes and Neutron Stars); DARK SECTOR (Dark Matter and Dark Energy); PLASMA ASTROPHYSICS (Origin of Jets, Cosmic Rays and Magnetic Fields) and the PRIMORDIAL UNIVERSE (Physics at the beginning of the Universe). In these four subjects, I discuss twelve of the most important challenges. These challenges give us insight into new physics that can only be studied in the large scale Universe. The near future possibilities, in observations and theory, for addressing these challenges, are also discussed.

------------------------------------------------------------------------

I. INTRODUCTION

There are many challenges in relativistic astrophysics. New satellites are being launched and new ground based telescopes are being inaugurated. Many theories are being investigated to explain the many phenomena. In the following sections I discuss what I feel are the most outstanding theoretical and observational challenges in relativistic astrophysics.

II. COMPACT OBJECTS (Black Holes and Neutron Stars)

CHALLENGE (1): Are there free quarks at the centers of neutron stars?

The equation-of-state at the center of neutron stars is a major challenge of relativistic astrophysics. Most models of dense matter predict that at densities greater than two times nuclear density at the centers of neutron stars, the formation of exotic matter takes place, such as free quarks (34). The gravitational wave signal from a binary containing a neutron star contains information on the equation-of-state of the neutron star, due to tidal polarizability of the neutron star (18, 19, 20, 21). Gravitational waves from newborn neutron stars, due to their oscillations, depend on the equation-of-state in the center. If the oscillation frequency of the emitted gravitational waves is ~ 2.6-3 kHz, a quark core is implied. The mass-radius relation of the neutron star also gives information on the equation-of-state of the neutron star. If there is a sufficiently high central pressure, implying a sufficiently high central density with a high kHz frequency, a quark core is indicated.

CHALLENGE (2): What are the characteristics of the gravitational waves emitted from merging black holes and neutron stars (number of polarization states; mass of graviton; speed of gravitational waves)?

The primary source of gravitational waves, for existing and near-future gravitational wave detectors, is merging binaries. Besides the emission of gravitational waves, the time evolution of merging binaries is affected by drag forces such as those due to magnetic fields. Due to a drag force, for example, relativity predicts an increase of the eccentricity of the orbits of a binary, opposite to the effect of gravitational wave emission (17). The number of polarization states of the gravitational waves may not be two, as predicted by General Relativity. Different theories predict a different number of polarization states. Other metric theories allow up to six states. Another interesting possibility of gravitational waves is that the graviton may have mass, which would make the speed of gravitational waves less than the speed of light.

CHALLENGE (3): Can the vacuum energy density become greater than other energy densities in the formation of relativistic stars?

In the absence of a full quantum gravity theory, the influence of gravity on quantum fields can be properly analyzed only in the semi-classical approximation, in which fields are quantized on a classical background space-time known as quantum field theory in curved space-time (QFTCS). Hawking radiation, the radiation of a black hole, is an example of this. It was shown that the vacuum energy density of a free quantum scalar field can become dominant over any classical energy-density. It was also shown that the natural time scale for this to happen is a tiny fraction of a second in the collapse of a relativistic star (40, 41). The formation and existence of neutron stars put limits on the possible contribution of the vacuum energy of the free quantum scalar field.

III. DARK SECTOR (Dark Matter and Dark Energy)

CHALLENGE (4): Where are the unseen thousands of predicted small dark matter halos, with central density cusps, in the Milky Way?

The standard LambdaCDM model of cosmology predicts thousands of small dark matter halos in the Milky Way, with central density cusps (density increasing as an inverse power of the radius, as the radius decreases). They could exist, without stars, with decreased baryon content, due to a primordial magnetic field preventing the entrance of baryons into the halos (13). A single supernova in the primordial universe could have expelled the baryons, which besides explaining their lack of stars, could have transformed the predicted central dark matter density cusp into a core (14). Gravitational lensing could detect these small dark matter halos and also determine if the central region is a cusp or core, but have not yet been able to do so.

CHALLENGE (5): What is the dark energy equation of state versus redshift?

One of the major astronomical observational efforts in the coming years is the determination of the dark energy equation of state versus redshift. The four most well established methods to determine the nature of dark energy are to observe: Type Ia supernovae; baryon acoustic oscillations (BAO); weak gravitational lensing; and the abundance of clusters of galaxies (12). Other approaches are observing: redshift-space distortions; the Alcock-Paczyinski effect; and direct measurements of the Hubble constant (12). If the equation of state is found observationally not to be constant, the deviation from a constant can give information on what is responsible for dark energy, such as an unknown field or modified gravity.

CHALLENGE (6): Can modified gravity explain dark energy?

Modified gravity could possibly explain dark energy. Experimental tests on Gpc scales rely on the growth and inter-relationship of perturbations in the metric potentials, and in the density and velocity fields, which can be measured using gravitational lensing, galaxy cluster abundances, galaxy clustering and the Sachs-Wolfe effect (33). Observations of modified gravity are interpreted in terms of the ratio of the two metric potentials where the ratio is unity in general relativity and non-zero in modified gravity. A strong indication against modified gravity would be if the ratio of the two metric potentials was found to be exactly unity.

IV. PLASMA ASTROPHYSICS (Origin of Jets, Cosmic Rays and Magnetic Fields)

CHALLENGE (7): What is the origin of relativistic jets?

Due to a magnetic field at the horizon of a spinning black hole, a relativistic jet can be produced (1, 2). The magnetic field at the horizon is a result of accretion or merging. In the merger of two black holes, there is an amplification of the magnetic field by two orders of magnitude (3). The high electrical conductivity of the outer layers of an accretion disk prevents the outward diffusion of a magnetic field, advecting the magnetic field into the horizon (4). It has been suggested that the interstellar magnetic field could be dragged inward and compressed by the accretion disk (28). Turbulence, however, in the disk could allow rapid diffusion of the field outward (28). The high electrical conductivity of the outer layers of the disk prevents the outward diffusion of the magnetic field implying a stationary state with zero radial velocity at the surface of the disk (30). General relativistic magnetohydrodynamic numerical calculations of the jet-accretion disk-compact object connection are now being done (31, 32). Detailed physics, however, of the accretion disk, in particular, its instabilities and nonlinear interactions and its connection with the jet and compact object, needs to be made. Relativistic jets could be hydrodynamically collimated (26) or magnetohydrodynamically collimated (27). Little work has been done on magnetohydrodynamic collimation since our 1990 study (27).

CHALLENGE (8): What is the origin of the ubiquitous microgauss magnetic fields in high and low redshift galaxies?

Almost all previous suggested origins are based on ad hoc assumptions that are difficult to prove or disprove. One of the most interesting suggestions for the origin of cosmic magnetic fields, which does not rely on any ad hoc assumption, is the natural fluctuations of the dense hot primordial plasma, as predicted by the Fluctuation Dissipation Theorem. It was shown that fluctuations produced in the high redshift plasma, after the primordial quark-hadron transition, could produce the present observed cosmic magnetic fields (7, 8). Many other scenarios have previously been suggested. We alone have suggested three other possible scenarios: (A) Non-minimal electromagnetic-gravitational coupling (i.e., rotation of a neutral body produces a magnetic field) (44, 45, 46); (B) Collimating magnetic fields of extragalactic jets spread out into the intergalactic medium (a magnetic field in the intergalactic medium will remain there undiminished for the age of the universe) (47); and (C) Density gradients not parallel to temperature gradients in primordial supernovae explosions (48).

CHALLENGE (9): What is the origin of the highest energy cosmic rays?

For the highest energy cosmic rays, we need an accelerator that is ten million times more powerful than the most powerful accelerator on Earth. The highest energy cosmic rays need to come from a distance less than 75 Mpc due to the GZK effect. The popular model for the source of high energy cosmic rays is First Order Fermi Acceleration. However, the strong shocks, associated with the few active galactic nuclei < 75 Mpc, such as Cen A, are not strongly correlated with these particles (22). Another possible source is the shocks in young supernovae remnants, where the ambient magnetic field has been amplified (23). Iron nuclei can be accelerated to a tenth of the highest energy cosmic rays in the observed amplified magnetic fields (24). The accelerated particles can slow down the incoming matter into the shock, eliminating entirely the observed shock (25). The accelerator also might be a spinning black hole in a strong magnetic field (2). No definite galactic source is known that can accelerate particles to ~3-5 X $10^{20}$ eV, the energy of the highest energy cosmic rays.

V. PRIMORDIAL UNIVERSE (Physics at the beginning of the Universe)

CHALLENGE (10): Did the Universe start to expand due to the very large initial vacuum Casimir energy?

The Casimir energy of the Universe is the vacuum energy due to its finite volume. Initially the Universe was very small and thus its Casimir energy was very large. We calculated the spatial distribution of the Casimir energy in the primordial Universe (42, 43). Cornish, Spergel and Starkman (49) suggested yhat a multiply connected "small" universe could allow for classical chaotic mixing as a preinflationary homogenization process. The smaller the volume, the more important the process. Also, a smaller universe has a greater probability of being spontaneously created from a quantum fluctuation. Dewitt, Hart and Isham (50) calculated the Casimir energy for static multiply connected flat space-times. A generalization of this calculation was made by us in (42, 43), showing that there is a spontaneous vacuum excitation of low multipolar components.

CHALLENGE (11): Is new physics needed to make a self-consistent theory of a primordial inflation era?

The assumption of having a primordial inflation period has had various successes. There are various problems, however. The energy scale of inflation is predicted to be on the order of a trillion times higher than the highest energy particle accelerators. In the past, each time when a higher energy particle accelerator was constructed, new physics was discovered. We may expect, therefore, that in the trillion energy interval there is new physics. In the popular chaotic inflation model of primordial inflation with a massive scalar field, the mass is unnaturally small. Another problem in the standard model of inflation is that the amplitude of the fluctuations created in the inflation era is dependent on the derivative of the potential with respect to the field, but the theory does not give its value. Still another problem is that the present popular model requires a very flat potential. However, no known particle has such a flat potential. A self-consistent theory of primordial inflation is thus still lacking. Inflation occurs at extremely high energies where quantum effects may occur, such as the non-commutation of space and time and a possible breakdown of Lorentz invariance. The non-commutation of space and time and violation of Lorentz invariance could possibly produce inflation (35, 36).

CHALLENGE (12): What is the precise value of the ratio, r, of primordial tensor-to-scalar fluctuations?

Different theories of the primordial Universe are sensitive to the ratio of primordial tensor-to-scalar fluctuations. Cosmic microwave background (CMB) astronomy is the primary means for obtaining information on the primordial universe. A measurement of the damping tail of the CMB with the South Pole Telescope plus the WMAP seven year release gives for the primordial tensor-to-scalar ratio r < 0.21 (9). There are no known technical limitations to achieving a sensitivity to detect r down to 0.001 (10). Backgrounds, however, will degrade this sensitivity, but a detection of r down to a level of 0.01 is achievable with a satellite mission (11). Due to the very long wavelengths of the tensor (gravitational wave) perturbations, their influence on structure formation is expected to be negligible.

VI. CONCLUSIONS

As discussed above, there are many exciting challenges in relativistic astrophysics. Better observational data and theoretical analysis in the coming years will help address these challenges.